\documentclass{article}
\usepackage{latexsym}
\usepackage{amsfonts}
\usepackage{amssymb}
\usepackage{amsmath}
\def\abstract#1{\vskip 7mm 
        \begin{center}{\large Abstract}\par \smallskip
                \begin{minipage}[c]{12cm}
                        \small #1
                \end{minipage}
        \end{center}
}
\def\title#1{\begin{center}{\Large\bf #1}\end{center}}
\def\author#1{\vskip 5mm \begin{center}{#1}\end{center}}
\def\address#1{\begin{center}{\it #1}\end{center}}
\makeatletter
\@ifundefined{lesssim}{}{}
\@ifundefined{gtrsim}{}{}
\def\vereq#1#2{\lower3pt\vbox{\baselineskip1.5pt \lineskip1.5pt
\ialign{$\m@th#1\hfill##\hfil$\crcr#2\crcr\sim\crcr}}}
\makeatother

\begin{document}

\title{%
 A Novel Approach to Braneworld
}
\author{%
  Jiro Soda \footnote{E-mail:jiro@tap.scphys.kyoto-u.ac.jp},
  Sugumi Kanno \footnote{E-mail:sugumi@tap.scphys.kyoto-u.ac.jp}
}
\address{%
   Department of Physics,  Kyoto University, Kyoto 606-8501, Japan
}

\abstract{
   Evaluating Kaluza-Klein (KK) corrections is indispensable to test the 
  braneworld scenario. In this report, we propose  a novel  symmetry
  approach to an effective 4-dimensional action with KK corrections
  for the Randall-Sundrum two-brane system.
}

\section{Introduction}

  It is generally believed that the singularity problem of the cosmology
  can be resolved in the context of the superstring theory. It seems
  that the most clear prediction of the superstring theory is the existence of
  the extra-dimensions. This apparently contradicts our experience.
  Fortunately, the superstring theory itself provides
  a mechanism to hide extra-dimensions, which is the so-called braneworld 
  scenario where the standard matter lives on the brane, 
  while only the gravity can feel the bulk space-time. In the seminal paper 
  by Randall and Sundrum, this scenario has been realized 
  in a two-brane model~\cite{RS1}. 
  Needless to say, it is important to test this new picture 
  by the cosmological observations in the context of this model. 
 
   As the observable quantities are usually represented by the 4-dimensional 
   language, it would be advantageous if we could find purely 4-dimensional
 description of the braneworld which includes the enough information of 
 the bulk  geometry, i.e. KK effects.  
 In the case of the single-brane model, 
 it is known that AdS/Conformal Field Theory (CFT) correspondence is a useful
 description of the braneworld~\cite{ads}. There, KK effects are interpreted
 as the contribution of CFT matter. 
 In the case of two-brane system, no such argument can be found 
  in the literature.
 What we need for the two-brane system 
 is a 4-dimensional description including KK effects 
 like the  AdS/CFT correspondence. 

 The purpose of this paper is  to present
  a novel approach that utilize the conformal symmetry as a
 principle  to determine the effective action. 
  Our new method not only
 gives a simple re-derivation of known results~\cite{KS1}, 
 but also leads to a new result, i.e. the effective action with KK
 corrections. 

The organization of this paper is as follows.
 In sec.II, we explain our method and re-derive known result.
 In sec.III, we derive new result, i.e. the KK corrected effective action.
 In the final section, we summarize our results and discuss
  possible applications and extension of our results. 
 Throughout this paper, we take the unit $8\pi G =1$.


\section{Symmetry Approach}
 
 For simplicity, we concentrate on the vacuum two-brane system. 
 Let us start with the 5-dimensional action  for this system
\begin{eqnarray}
  S [\gamma_{AB}, g_{\mu\nu}, h_{\mu\nu}] 
\end{eqnarray}
where $\gamma_{AB}$, $g_{\mu\nu}$ and $h_{\mu\nu}$ are the 5-dimensional
 bulk metric, the induced metric on the positive and the negative 
 tension branes, respectively.    
 Now,  suppose to solve the bulk equations of motion and the 
 junction condition on the negative tension brane, 
 then formally we get the relation
\begin{eqnarray}
  \gamma_{AB} = \gamma_{AB}[g_{\mu\nu}] 
  \ , \quad h_{\mu\nu} = h_{\mu\nu} [g_{\mu\nu}] \ .
  \label{bulk-metric}
\end{eqnarray}
By substituting relations (\ref{bulk-metric})  into the original action,
 in principle, the 4-dimensional effective action can be obtained as
\begin{eqnarray}
 S_{\rm eff} 
 = S[\gamma_{AB} [g_{\mu\nu}] ,g_{\mu\nu}, h_{\mu\nu}[g_{\mu\nu}]] \ .
 \label{eff-action}
\end{eqnarray}
 Unfortunately, the above calculation is not feasible in practice. 
 In the following, we propose a novel method to deduce the effective action.
 
 Let us take the gradient expansion approach at the action level.
 At low energy,   it seems legitimate to assume that 
 the action can be expanded by the local terms with increasing
  orders of derivatives if one includes all of the relevant 
 degrees of freedom~\cite{KS1}. 
 In the two-brane system, the relevant degrees of freedom
 are nothing but the metric  and  the radion  which can be seen from the linear
 analysis~\cite{GT}. 
  Hence, we assume the general local action constructed from the 
  metric $g_{\mu\nu}$ and the radion $\Psi $ as an ansatz. 
 Therefore,  we can write the action as 
\begin{eqnarray}
S_{\rm eff}
    &=&
	{1\over 2} \int d^4 x \sqrt{-g} \left[ \Psi R - 2\Lambda (\Psi )
	-{\omega (\Psi) \over \Psi} \nabla^\mu \Psi \nabla_\mu \Psi \right]
        \nonumber\\
&&\!\!
	+\int d^4 x \sqrt{-g} \left[
	A(\Psi) \left( \nabla^\mu \Psi \nabla_\mu \Psi \right)^2
	+B(\Psi) \left( \Box \Psi \right)^2 
	+C(\Psi)  \nabla^\mu \Psi \nabla_\mu \Psi \Box \Psi
	+D(\Psi) R~\Box \Psi \right. \nonumber\\
&&\left.\!\!\!\!  
	+ E(\Psi) R \nabla^\mu \Psi \nabla_\mu \Psi
     + F(\Psi) R^{\mu\nu} \nabla_\mu \Psi \nabla_\nu \Psi 
     + G(\Psi) R^2     
	+ H(\Psi) R^{\mu\nu} R_{\mu\nu} 
	+I(\Psi) R^{\mu\nu\lambda\rho} R_{\mu\nu\lambda\rho} 
	+\cdots  \right]  \ , 
	\label{setup}
\end{eqnarray}
where we have listed up all of the possible local terms 
which have derivatives up to fourth-order. This series will continue infinitely.
 We have the freedom to redefine the scalar field $\Psi$. In fact, we have used
 this freedom to fix the functional form of the coefficient of $R$. 
 To determine other coefficient functions, 
  we use  the geometric method which yields, 
 instead of the action,  directly 
 the effective equations of motion~\cite{ShiMaSa} 
\begin{eqnarray}
  G_{\mu\nu} = T_{\mu\nu} + \pi_{\mu\nu} - E_{\mu\nu}  
\end{eqnarray}
where $T_{\mu\nu}$ is the energy-momentum tensor of the matter and
\begin{eqnarray}
  \pi_{\mu\nu} = -{1\over 4} T_{\mu\lambda} T^\lambda{}_{\nu}
      + {1\over 12} T T_{\mu\nu} 
    +{1\over 8}g_{\mu\nu} \left( T^{\alpha\beta} T_{\alpha\beta} 
      -{1\over 3} T^2 \right) \ . 
\end{eqnarray}
Note that the projection of Weyl tensor $E_{\mu\nu}$ represents  the
 effect of the bulk geometry. 
For the vacuum brane which we are considering, this reduces to 
\begin{eqnarray}
   G_{\mu\nu} =  - E_{\mu\nu} - \lambda g_{\mu\nu}   \ ,
\end{eqnarray}
where we have renormalized the cosmological constant  $\lambda$  
so that it includes the quadratic part of the energy-momentum tensor. 
One defect of this approach is that $E_{\mu\nu}$ is not known
 except for the following property
\begin{eqnarray}
   E^\mu{}_\mu =0 \ .
   \label{traceless}
\end{eqnarray}
 For the isotropic homogeneous universe, Eq.~(\ref{traceless}) 
 has sufficient information to deduce the cosmological evolution equation.
  For general spacetimes, however, this traceless condition is not sufficient
  to determine the evolution of the braneworld. 
 However, combination of the geometric approach and the gradient expansion 
 approach determines the effective action. Now, we  explain our method.
  We have introduced the radion explicitly in the gradient expansion approach. 
 While the radion never appears in the geometric approach, 
 instead  $E_{\mu\nu}$ 
 is induced as the effective energy-momentum tensor reflecting the
 effects of the bulk geometry.  
 Notice that the  property (\ref{traceless}) implies the conformal invariance 
 of this effective matter. Clearly, both approaches should agree to each other.
  Hence, the radion  must play a role of the conformally invariant matter 
 $E_{\mu\nu}$.   This symmetry requirement gives a stringent constraint on the
 action, more precisely,  the conformal symmetry (\ref{traceless}) determines
 radion dependent  coefficients  in the action (\ref{setup}). 

 Let us illustrate our method using the following truncated action 
\begin{eqnarray}
    S_{\rm eff} 
    ={1\over 2}  \int d^4 x \sqrt{-g} \left[ 
          \Psi R -2\Lambda (\Psi) - {\omega (\Psi) \over \Psi} 
          \nabla^\mu \Psi \nabla_\mu \Psi \right] \ ,
\end{eqnarray}
 which is nothing but the scalar-tensor theory with coupling function
 $\omega (\Psi)$ and the potential function $\Lambda (\Psi )$. 
 Note that this is the most general local action which contains 
 up to the second  order derivatives and has the general coordinate invariance.
 First, we must find $E_{\mu\nu}$. 
The above action gives the equations of motion for the metric as
\begin{eqnarray}
    G_{\mu\nu} = -{\Lambda \over \Psi} g_{\mu\nu}
                   + {1\over \Psi} \left( 
                 \nabla_\mu \nabla_\nu \Psi - g_{\mu\nu} \Box \Psi \right)
                      + {\omega \over \Psi^2} \left(
             \nabla_\mu \Psi \nabla_\nu \Psi  -{1\over 2} g_{\mu\nu}
             \nabla^\alpha \Psi \nabla_\alpha \Psi \right)  \ . 
\end{eqnarray}
The right hand side of this Eq.~(11) should be identified with 
$-E_{\mu\nu}-\lambda g_{\mu\nu}$.
 Hence, the  condition  $E^\mu{}_\mu =0$ becomes
\begin{eqnarray}
        \Box \Psi = - {\omega \over 3\Psi} 
        \nabla^\mu \Psi \nabla_\mu \Psi  
        + {4\over 3} \left( \Lambda  - \lambda \Psi \right)  \ .
\end{eqnarray}
This is the equation for the radion $\Psi$. However, we also
have the equation for $\Psi$ from the action as 
\begin{eqnarray}
    \Box \Psi = \left( {1\over 2\Psi} - { \omega' \over 2\omega} \right)
             \nabla^\alpha \Psi \nabla_\alpha \Psi  
             + {\Psi \over 2\omega} R - {\Psi \over \omega} \Lambda'    \ ,
\end{eqnarray}
where the prime denotes the derivative with respect to $\Psi$. 
In order for these two Eqs.~(12) and (13) to be compatible, $\Lambda$ and
  $\omega$ must satisfy 
\begin{eqnarray}
  {1\over 2\Psi} - { \omega' \over 2\omega} 
	=-{\omega \over 3 \Psi} \ ,   \quad
  {4\over 3} \left( \Lambda - \lambda \Psi \right) =
            {\Psi \over \omega} \left( 2\lambda - \Lambda' \right)  \ ,
\end{eqnarray}
where we used  $R= 4\lambda$ which comes from the trace part of Eq.~(7).
 Eqs.~(14) and (15) can be integrated as  
\begin{eqnarray}
   \Lambda (\Psi) = \lambda + \lambda \beta \left( 1-\Psi \right)^2   \ , \quad
  \omega (\Psi ) = {3\over 2} {\Psi \over 1-\Psi}  \ ,  
\end{eqnarray}
where the constant of integration $\beta$ represents the ratio
 of the cosmological constant on the negative tension brane to that on 
 the positive tension brane~\cite{KS1}. 
 Here, one of constants of integration is absorbed by rescaling of $\Psi$.
 In doing so, we have assumed the constant of integration is positive.
 The case of negative signature corresponds to the negative tension brane.
 In other words, we can also describe the negative tension brane 
 in this method.

Thus, we get the effective action 
\begin{eqnarray}
S_{\rm eff}
	= \int d^4 x \sqrt{-g} \left[ {1\over 2} \Psi R 
	-{3 \over 4( 1-\Psi )} \nabla^\mu \Psi \nabla_\mu \Psi 
    - \lambda - \lambda \beta (1-\Psi)^2 \right] \ .
\end{eqnarray}
This completely agrees with the previous 
result~\cite{KS1}.     Surprinsingly, our simple symmetry 
 approach has determined the action completely.


\section{KK corrections}

  Let us extend the result in the previous section
 to the higher order case. 
 We have already determined the functions $\Lambda (\Psi)$ and $\omega (\Psi )$.
 From the linear analysis, the action in the previous section is known 
 to come only from zero modes. Hence, one can expect the other coefficients 
 in the action (\ref{setup}) represent the effects of KK-modes. 
 
  Now we impose the conformal symmetry  on the fourth order derivative terms
  in the action (\ref{setup}) as we did in the  previous section. 
  Starting from the action (\ref{setup}), one can read off the equation for 
  the metric and hence $E_{\mu\nu}$ can be identified. 
 The compatibility condition between $E^\mu{}_\mu =0$ and the equation for the
 radion $\Psi$ leads to a set of equations which seems to be over constrained.
  Nevertheless, one can find solutions consistently. 
 Thus, we find the 4-dimensional effective action with KK corrections 
 as~\cite{Kanno}
\begin{eqnarray}
S_{\rm eff} 
  &=& \int d^4 x \sqrt{-g} \left[ {1\over 2} \Psi R 
     - {3 \over 4( 1-\Psi )} \nabla^\mu \Psi \nabla_\mu \Psi 
    - \lambda - \lambda \beta (1-\Psi)^2    \right]
              \nonumber\\
 &&    + \ell^2 \int d^4 x \sqrt{-g} \left[
     {1 \over 4 (1-\Psi)^4} \left( \nabla^\mu \Psi \nabla_\mu \Psi \right)^2
         + {1\over (1-\Psi)^2} \left( \Box \Psi \right)^2 
      + {1\over (1-\Psi)^3} \nabla^\mu \Psi \nabla_\mu \Psi\Box \Psi
      \right. \nonumber \\
 && \left.   + {2\over 3(1-\Psi)} R \Box \Psi 
      + {1\over 3(1-\Psi)^2}  R \nabla^\mu \Psi \nabla_\mu \Psi 
       + g R^2     + h R^{\mu\nu} R_{\mu\nu}  \right] \ .
 \label{action}
\end{eqnarray}
 Because of the existence of the Gauss-Bonnet topological term, we can omit
 the square of Riemann tensor without losing the generality. 
 The constants $g$ and $h$ can be interpreted as 
 the variety of effects of the bulk gravitational field.


\section{Conclusion}

  We have established  a novel symmetry 
  approach to  an effective  4-dimensional action with KK corrections. 
  This is done by combining the low energy expansion of the action
  and the geometric approach. Our result supports the smoothness
  of the collision process of two branes advocated 
  in the ekpyrotic (cyclic) model and  born-again model. 
  Not only our result can be used to assess the validity of
 the low energy approximation, but also has a potential to make concrete
 predictions to be compared with observations. 

 As to the cosmological applications, it is important to recognize
 that our action can describe the inflation.  
 Cosmological perturbations~\cite{soda} are
 now ready to be studied. 
 In fact, our result provides the basis
 of the prediction of CMB spectrum with KK corrections.
 
 The black hole solutions with KK corrections
 are also interesting subjects. 
 If we truncate the system at the lowest order, the static solution 
 is Schwartzshild black hole with a trivial radion which corresponds 
 to the black string in the bulk. 
 The Gregorry-Laflamme instability occurs when the wavelength of KK modes
 exceeds the gravitational length of the black hole~\cite{GL}. 
 Clearly, the lightest
 KK mode is important and this mode is already included in our action, 
 hence it would be  interesting to investigate if 
  the Gregory-Laflamme instability occurs or not within our 
  theory~\cite{tamaki,KS2}.

\vskip 0.5cm 
\noindent{Acknowledgements}\\
\smallskip
 The authors would like to thank Kei-ichi Maeda for useful comments. 
This work was supported in part by  Grant-in-Aid for  Scientific
Research Fund of the Ministry of Education, Science and Culture of Japan 
 No. 155476 (SK) and  No.14540258 (JS).

\end{document}